\documentclass[preprint,superscriptaddress,floatfix]{revtex4}
\usepackage{graphicx}
\usepackage{amssymb,amsmath,latexsym}
\usepackage{bm}
\usepackage[dvips]{epsfig}
\usepackage{mathptmx}

\newcommand{\be}{\begin{equation}}
\newcommand{\ee}{\end{equation}}
\newcommand{\ba}{\begin{eqnarray}}
\newcommand{\ea}{\end{eqnarray}}
\newcommand{\parallelsum}{\mathbin{\!/\mkern-5mu/\!}}

\makeatletter
\def\@seccntformat#1{%
  \expandafter\ifx\csname c@#1\endcsname\c@section\else
  \csname the#1\endcsname\quad
  \fi}
\makeatother

\begin{document}
\title{Spin-wave propagation in cubic anisotropic materials}
\author{Koji Sekiguchi}%
\thanks{These two authors contributed equally to this work.}
\affiliation{Department of Physics, Keio University, Hiyoshi 3-14-1, Yokohama 223-8522, Japan}%
\affiliation{JST-PRESTO, Gobanchon 7, Chiyoda-ku, Tokyo 102-0076, Japan}%
\author{Seo-Won Lee}%
\thanks{These two authors contributed equally to this work.}
\affiliation{Department of Materials Science and Engineering, Korea
University, Seoul 02841, Korea}%
\author{Hiroaki Sukegawa}%
\affiliation{National Institute for Materials Science (NIMS), 1-2-1 Sengen, Tsukuba 305-0047, Japan}%
\author{Nana~Sato}%
\affiliation{Department of Physics, Keio University, Hiyoshi 3-14-1, Yokohama 223-8522, Japan}%
\author{Se-Hyeok Oh}%
\affiliation{Department of Nano-Semiconductor and Engineering, Korea University, Seoul 02841, Korea}%
\author{Robert D. McMichael}%
\affiliation{Center for Nanoscale Science and Technology, National Institute
of Standards and Technology, Gaithersburg, Maryland 20899, USA}%
\author{Kyung-Jin Lee}%
\email{kj_lee@korea.ac.kr}%
\affiliation{Department of Materials Science and Engineering, Korea
University, Seoul 02841, Korea}%
\affiliation{Department of Nano-Semiconductor and Engineering, Korea University, Seoul 02841, Korea}%
\affiliation{KU-KIST Graduate School of Converging Science and Technology, Korea University, Seoul 02841, Korea}%
%
%
%
\begin{abstract}
The information carrier of modern technologies is the electron charge whose transport inevitably generates Joule heating. Spin-waves, the collective precessional motion of electron spins, do not involve moving charges and thus avoid Joule heating. In this respect, magnonic devices in which the information is carried by spin-waves attract interest for low-power computing. However implementation of magnonic devices for practical use suffers from low spin-wave signal and on/off ratio. Here we demonstrate that cubic anisotropic materials can enhance spin-wave signals by improving spin-wave amplitude as well as group velocity and attenuation length. Furthermore, cubic anisotropic material shows an enhanced on/off ratio through a laterally localized edge mode, which closely mimics the gate-controlled conducting channel in traditional field-effect transistors. These attractive features of cubic anisotropic materials will invigorate magnonics research towards wave-based functional devices.
\end{abstract}
%
%
\maketitle

\section{INTRODUCTION}
\vspace{-3mm}
\noindent
Magnonics is a research field that aims to control and manipulate spin-waves in magnetic materials for information processing~\cite{Serga1, Kruglyak, Lenk}. Spin-waves enable Boolean and Non-Boolean computing with low-power consumption ~\cite{Khitun1, Schneider, Lee, Khitun2, Csaba, Sato, Chumak1}. Their wave properties also allow distinct functionalities~\cite{Vogt2, Vogel, Haldar,Kwon} such as multi-input/output (non-linear) operations~\cite{Khitun3,Khitun4}. Despite significant progress, however, low signal and on/off ratio of spin-waves have been major obstacles to implementation of magnonic devices for practical use. The on/off ratio of spin wave can be related to the asymmetric ratio of spin wave amplitude. These obstacles are caused by poor excitation efficiency and propagation losses of spin-waves. In this work, we show that cubic anisotropic materials offer a more efficient spin-wave propagation than conventional materials.

Our strategy to improve the spin-wave signal is to modify the dispersion relation. Compared to other waves in solid states, the spin-wave dispersion is highly anisotropic, caused by the long-range magnetostatic interaction. An established way to control the anisotropy of dispersion is to change the relative orientation between the equilibrium magnetization direction $\textbf{m}$ and the direction of wave vector $\textbf{k}$ [i.e., backward volume ($\textbf{m} \parallelsum \textbf{k}$), surface ($\textbf{m} \bot \textbf{k}$ for in-plane $\textbf{m}$), and forward volume modes ($\textbf{m} \bot \textbf{k}$ for out-of-plane $\textbf{m}$)]. As the dispersion determines all spin-wave properties, a proper modification of the dispersion may allow us to improve spin-wave properties. We note that for improved functionalities of spin-wave devices, not only the spin-wave amplitude but also the spin-wave attenuation length and group velocity should be improved.

Here we introduce an epitaxial Fe film as a waveguide for this purpose. The cubic crystalline anisotropy of epitaxial Fe film provides an additional knob to modify the dispersion by changing the relative orientation between the magnetic easy (or hard) axis and the direction of wave vector $\textbf{k}$. As we show below, a proper tuning of this relative orientation allows an enhancement of the spin-wave amplitude by a factor of 28, as well as enhancements of the spin-wave attenuation length and group velocity by several factors. We also show that the cubic anisotropy provides a giant lateral spin-wave asymmetry due to a laterally localized edge mode at zero external field, which may enable three-terminal nonvolatile spin-wave logic gates.

\section{Materials and Theoretical framework}
\vspace{-3mm}
\noindent
The spin-wave dispersion in cubic anisotropic materials was first derived by Kalinikos {\it et al.}~\cite{Kalinikos}. Instead of starting with this known dispersion directly, here we describe the problem in a rather general way in order to get an insight of how to improve spin-wave properties. We first describe two key requirements in the dispersion relation for improved spin-wave properties. From the spin-wave theory for an in-plane magnetization ${\bf m}$ with an assumption that $\bf m$ varies only along the spin-wave propagation direction~\cite{DE, Patton, KS, Bob}, the spin-wave amplitude $A_{\rm SW}$, the group velocity $v_{\rm g}$, and the attenuation length $\Lambda$ are respectively given as (see Supplementary Note 1 for details),
\begin{eqnarray}
A_{\rm SW} &\propto& \sqrt{\frac{H_2}{H_1}}\frac{M_{\rm s}}{\alpha_{\rm G}(H_1+H_2)}, \\
v_{\rm g} &=& \frac{\gamma_{\rm g} \mu_0 M_{\rm s}}{2}\frac{\partial P_k}{\partial k}\left(\sqrt{\frac{H_2}{H_1}}\sin^2\phi-\sqrt{\frac{H_1}{H_2}}\right), \\
\Lambda &=& \frac{M_{\rm s}}{\alpha_{\rm G} (H_1+H_2)}\frac{\partial P_k}{\partial k}\left(\sqrt{\frac{H_2}{H_1}}\sin^2\phi-\sqrt{\frac{H_1}{H_2}}\right),
\end{eqnarray}
where $H_1$ (=$F_{\phi\phi}/\mu_0 M_{\rm s}+M_{\rm s}P_k\sin^2\phi$) and $H_2$ (=$F_{\theta\theta}/\mu_0 M_{\rm s}+M_{\rm s}(1-P_k)$) are the in-plane and normal effective fields, respectively and $F$ is the free magnetic energy density without magnetostatic interactions, which are treated separately.  The subscripts (i.e., $\theta$ and $\phi$ are the polar and azimuthal angles of ${\bf m}$, respectively) on $F$ refer to partial derivatives around equilibrium positions, $P_k=1-(1-{\rm e}^{-|k|d})/|k|d$, $k$ is the wavenumber, $d$ is the film thickness, $M_{\rm s}$ is the saturation magnetization, $\alpha_{\rm G}$ is the Gilbert damping, and $\gamma_{\rm g}$ is the gyromagnetic ratio. In Eq. (1), we refer to the dominant in-plane component only because the normal component is negligible due to strong demagnetization of thin film geometry.

Focusing on the limit of $kd \ll 1$ with which $P_k$ is small, one finds from Eqs. (1-3) that $A_{\rm SW}$, $v_{\rm g}$, and $\Lambda$ are simultaneously maximized when $H_1 \approx 0$ (equivalently $F_{\phi\phi} \approx 0$) and $\phi \approx \pm\pi/2$ (i.e., surface mode configuration). Therefore, two key requirements for the improved spin-wave properties are {\it vanishingly small in-plane effective field and surface mode}. For $kd \ll 1$, $F$ consists of the Zeeman and anisotropy energies. The contribution from the Zeeman energy to $F_{\phi\phi}$ is simply an external in-plane field $H (>0)$ that should be applied to ensure the surface mode. The central question is thus how to obtain a negative in-plane effective field from the anisotropy energy in order to diminish $F_{\phi\phi}$ in the surface mode.

\begin{figure}[ttbp]
\begin{center}
\psfig{file=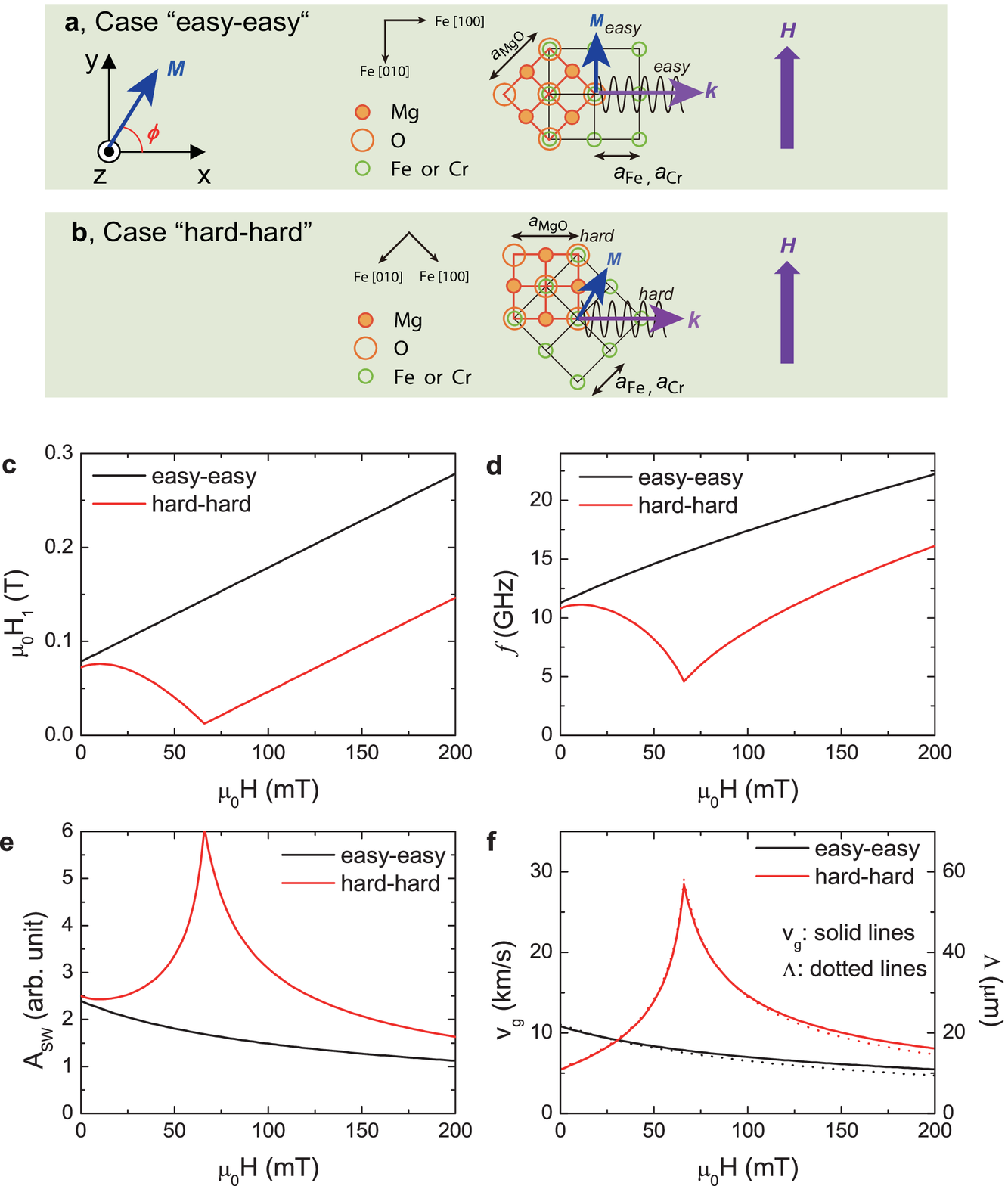,width=0.7\columnwidth} \caption{\label{fg1}
{\bf One-dimensional theoretical calculations of spin-wave properties in cubic anisotropy material.} Top view of planar system of {\bf a}, {\it easy-easy} case and {\bf b}, {\it hard-hard} case where the crystallographic orientations are described for an epitaxial Fe layer with a cubic crystalline anisotropy. The long axis of the waveguide (i.e., direction of wave vector $\textbf{k}$) for the {\it easy-easy} device is along the easy axis (Fe [100] direction, {\bf a}) whereas that for the {\it hard-hard} device is along the hard axis (Fe [1${\bar 1}$0] direction, {\bf b}). Computed  results based on spin-wave theories [Eqs. (1-3)]. {\bf c}, In-plane effective field $H_1$ versus external field $H$. {\bf d}, Spin-wave frequency $f$ versus $H$. {\bf e}, Spin-wave amplitude $A_{\rm SW}$ versus $H$. {\bf f}, Group velocity $v_{\rm g}$ and attenuation length $\Lambda$ versus $H$. Parameters: $M_s$=1600 kA/m, $\mu_0 H_A$=66 mT, $k=5 \times 10^5$ m$^{-1}$, $\gamma_g = 1.76 \times 10^{11}$ T$^{-1}$s$^{-1}$, $\alpha$=0.0026, and $d$=25 nm. }
\end{center}
\end{figure}

We next show that spin-wave propagation where the wave vector $\textbf{k}$ is along the hard-axis direction of cubic anisotropic materials can naturally satisfy these two requirements. We consider two cases, {\it easy-easy} (Fig.~\ref{fg1}a) and {\it hard-hard} (Fig.~\ref{fg1}b) cases, in which the first (second) word corresponds to the direction of ${\bf m}$ (wave vector $\textbf{k}$). A top view of waveguide for each case is shown in Fig. ~\ref{fg1}a and ~\ref{fg1}b where the crystallographic orientations are defined for an epitaxial Fe layer with a cubic crystalline anisotropy ~\cite{Sukegawa}. The long axis of the waveguide (i.e., direction of wave vector $\textbf{k}$) for the {\it easy-easy} device is along the easy axis (Fe [100] direction, Fig. ~\ref{fg1}a) whereas that for the {\it hard-hard} device is along the hard axis (Fe [1${\bar 1}$0] direction, Fig. ~\ref{fg1}b). Our main focus is the {\it hard-hard} case, whereas the {\it easy-easy} case corresponds to conventional spin-wave propagation and will be used as a reference. In both cases, the external field $H$ is applied in the $y$-direction and wave vector ($\textbf{k}$) is in the $x$-direction. The cubic anisotropy energy density $E_{an}$ is given as
\be\label{Ean}
E_{an}=K_c (\cos^2 \alpha \cos^2 \beta+\cos^2 \beta \cos^2 \gamma+\cos^2 \gamma \cos^2 \alpha),
\ee
where $K_c$ is the cubic anisotropy and $\alpha (=\phi-\phi_K)$, $\beta(=\pi/2-\phi+\phi_K)$, and $\gamma(=\theta=\pi/2)$ are direction cosines of the magnetization with respect to the easy axes of cubic anisotropy. Here $\phi_K$ is the angle between the easy axis and $x$-axis, and is 0 ($\pi/4)$ for the {\it easy-easy} ({\it hard-hard}) case. On the other hand, the Zeeman energy is $-\mu_0 M_s H \sin\phi \sin\theta$. With these energy terms, $H_1$ for the {\it hard-hard} and {\it easy-easy} cases are readily calculated as
\begin{eqnarray}
H_1^{\rm hard} &=& H\sin\phi_{\rm eq}-H_{\rm A} \cos(4\phi_{\rm eq})+M_{\rm s} P_k \sin^2\phi_{\rm eq}, \\
H_1^{\rm easy} &=& H+H_{\rm A}+M_{\rm s} P_k,
\end{eqnarray}
where $H_{\rm A} (=2K_c/\mu_0 M_s)$ is the cubic anisotropy field, $\phi_{\rm eq} = \pi/2$ for $h \ge 1$, $h=H/H_{\rm A}$, and

\begin{equation}
 \phi_{\rm eq} = \sin^{-1} \left(\frac{6^{1/3}+(9h+\sqrt{81h^2-6})^{2/3}}{6^{2/3}(9h+\sqrt{81h^2-6})^{1/3}} \right),
\end{equation}
for $0 \le h<1$.

For $h \ge 1$, $H_1^{\rm hard}$ becomes $H-H_{\rm A}+M_{\rm s} P_k$. For the {\it hard-hard} case, therefore, the cubic anisotropy provides a negative effective field and $H_1^{\rm hard}$ becomes small when $H \approx H_{\rm A}$ and $\phi_{\rm eq}=\pi/2$ because $M_{\rm s} P_k$ is small. Figure~\ref{fg1}c shows that this is indeed the case. For the {\it hard-hard} case, the frequency $f(=\omega/2\pi)$ minimizes at $H \approx H_{\rm A}$ (Fig.~\ref{fg1}d) and all of $A_{\rm SW}$, $v_{\rm g}$, and $\Lambda$ are largely enhanced at $H \approx H_{\rm A}$, in comparison for the {\it easy-easy} case (Fig.~\ref{fg1}e and \ref{fg1}f). We note that the improvements of spin-wave properties discussed in this section are consequences of the known dispersion~\cite{Kalinikos}, but there has been no direct experimental proof for spin-wave waveguides made of cubic anisotropic materials.


\section{Results and discussion}
\begin{figure*}[bhp]
\begin{center}
\psfig{file=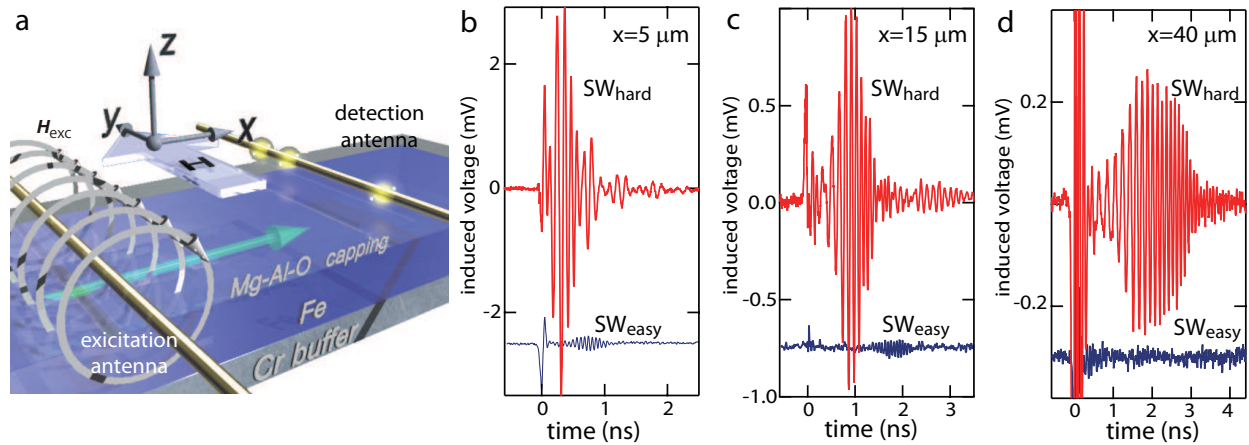,width=1.0\columnwidth}  \caption{\label{figure2}
{\bf Large-amplitude propagating spin waves in an epitaxial Fe waveguide.} {\bf a}, Schematic illustration of the time-domain propagating spin-wave spectroscopy. Spin-wave packets at distances {\bf b}, 5 $\mu$m, {\bf c}, 15 $\mu$m, and {\bf d}, 40 $\mu$m. The external magnetic field is 70 mT. At time $t$=0, a pulsed voltage is applied to excite spin-waves. In {\bf b}-{\bf d}, the curves are intentionally offset for clarity.
  }
\end{center}
\end{figure*}

\vspace{-3mm}
\noindent
In order to confirm the theoretical prediction, we measure spin-wave-induced voltages in the time domain for microfabricated devices containing two antennas [i.e., spin-wave excitation and detection antennas (Fig. \ref{figure2}a), see Methods], based on the propagating spin-wave spectroscopy~\cite{Covington, Sekiguchi1}. The layer structure of the waveguide is Cr(40)/Fe(25)/Mg(0.45)/Mg-Al(1.2)/oxidation (thicknesses in nanometers; see Methods), where the epitaxial Fe layer has cubic crystalline anisotropy~\cite{Sukegawa}. $\mu_0 H_{\rm A}$ of the Fe film is (66 $\pm$2) mT, determined by the vibrating sample magnetometer (Supplementary Note 2). We fabricate spin-wave devices with varying antenna distances in order to measure the group velocity and attenuation length.

Figure \ref{figure2}b-d show representative time-domain results at $H \approx H_{\rm A}$ and various antenna distances. It clearly shows that the spin-wave amplitude is much larger for the {\it hard-hard} case than for the {\it easy-easy} case. Magnetic field $H$-dependences of $A_{\rm SW}$, $v_{\rm g}$, and $\Lambda$ are summarized in Fig. \ref{figure3}. All the experimental results show enhanced spin-wave properties at $H \approx H_{\rm A}$ as predicted by the theory.
At $H \approx H_{\rm A}$ and the antenna distance of 5 $\mu$m, the induced voltage of spin-wave packet reaches 3.30 mV for the {\it hard-hard} case (Fig. \ref{figure3}a), whereas it is about 0.12 mV for the {\it easy-easy} case (Fig. \ref{figure3}c). Therefore, the spin-wave amplitude $A_{\rm SW}$ for the {\it hard-hard} case enhances by a factor of 28 in comparison for the {\it easy-easy} case. By analyzing the antenna-distance dependence of amplitudes and arrival-times of spin-wave packets~\cite{Sekiguchi1}, we deduce the spin-wave attenuation length $\Lambda $ and the spin-wave group velocity $v_{\rm g}$. At the enhancement condition ($\mu_0 H=$ 66 mT), $\Lambda$ for the \textit{hard-hard} case is about (17.8 $\pm$ 0.5) $\mu$m whereas $\Lambda $ for the \textit{easy-easy} case remains (10.0 $\pm$ 4.5) $\mu$m (Fig. \ref{figure3}f). At the enhancement condition, furthermore, $v_{\rm g}$ for the \textit{hard-hard} case is about (23.4 $\pm$ 0.7) km/s whereas $v_g$ for the \textit{easy-easy} case is about (8.9 $\pm$ 0.3) km/s (Fig. \ref{figure3}g). We note that uncertainties expressed in this paper are one standard deviation of fit parameters.

Therefore, all these results confirm that spin-wave properties are largely improved at the enhancement condition, qualitatively consistent with the theoretical predictions. For the {\it hard-hard} case, we find however that there are interesting quantitative differences between the experimental and theoretical ones. For instance, the one-dimensional theory (Fig. \ref{fg1}) predicts that the ratio of $A_{\rm SW}$ at $H = H_{\rm A}$ to $A_{\rm SW}$ at $H = 0$ is about 3.5 with considering spin-wave attenuation for the distance of 5 $\mu$m. This predicted ratio is much smaller than the experimental one ($\approx$ 9;  $A_{\rm SW}^{H = 0}$ $\approx$ 0.36 mV and $A_{\rm SW}^{H = H_{\rm A}}$ $\approx$ 3.30 mV, see Fig. \ref{figure3}a).

In order to understand this discrepancy, we perform two-dimensional micromagnetic simulations for the {\it hard-hard} case (Fig. \ref{figure4}). When $\textbf{m}$ is aligned along one of easy-axes (i.e., $H=0$ and $\phi=\pi/4$; Fig. \ref{figure4}a), the spin-wave having $-k$ ($+k$) is localized at the top-left (bottom-right) edge. The exact opposite trend is obtained when $\textbf{m}$ is aligned along another easy-axis (i.e, $H=0$ and $\phi=3\pi/4$; Fig. \ref{figure4}d). On the other hand, this laterally localized spin-wave edge mode is absent when $H = H_{\rm A}$ (Fig. \ref{figure4}b). This edge mode originates from the anisotropic dispersion relation depicted in Fig. \ref{figure4}c. It shows a contour plot of the spin wave dispersion relation of hard-hard case at $H=0$. The dispersion relation at the excitation frequency ($10.5$ GHz) is highlighted. One finds that the wave fronts propagate in the direction of $\textbf{k}$ determined primarily by the antenna geometry whereas the group velocity, $v_g={\rm d}\omega/{\rm d}k$, is perpendicular to the contour, the oblique direction.
This anisotropic propagation leads to the edge localization of spin-waves, which in turn makes an additional difference of $A_{\rm SW}$ between $H = H_{\rm A}$ and $H = 0$ because our experimental set-up detects an induced voltage integrated over the full width of waveguide.

This localized edge mode at zero field allows us to significantly improve spin-wave asymmetry using cubic anisotropic materials. It can be compared to the conventional spin-wave nonreciprocity in the aspect of ``asymmetric propagation of spin-wave''. Conventional spin-wave nonreciprocity refers to an asymmetric spin-wave amplitude depending on the spin-wave propagation direction when spin-waves are excited by a magnetic field generated by microwave antennas~\cite{Bailleul, Demidov, Sekiguchi2, Jamali}. This amplitude asymmetry results from a nonreciprocal antenna-spin-wave coupling, caused by the spatially nonuniform distribution of the antenna field. The asymmetry factor ($\equiv A_{\rm SW}^{{\bf k}>0}/A_{\rm SW}^{{\bf k}<0}$) for the conventional spin-wave nonreciprocity is about 2~\cite{Sekiguchi2, Jamali}. In contrast, the lateral spin-wave asymmetry in cubic anisotropic materials can be very large when the detection antenna is properly designed. Figure \ref{figure4}e shows that the spin-wave profile with respect to the location of the excitation antenna highly is symmetric at the top and bottom edges. By placing a detection antenna at one of the edges (see Fig. \ref{figure4}d), therefore, one can obtain a very large difference in the spin-wave amplitude between the cases for $\phi=\pi/4$ and for $\phi=3\pi/4$ (the asymmetry factor $\approx$ 40 for Fig. \ref{figure4}f).

In order to confirm the lateral spin-wave asymmetry, we experimentally measure the spatial distribution of the magnon density by micro-focused Brillouin Light Scattering (BLS) spectroscopy~\cite{DemokritovBLS} (see Methods) for the {\it hard-hard} device. We inject an RF current at $10.4$ GHz in the excitation antenna and measure the BLS spectra at $H=0$ as a function of the distance $x$ from the signal line (Fig. \ref{figure5}a). The BLS spectra at an edge (i.e., bottom edge, $y$ = 115 $\mu$m) show a clear dependence on $x$ (Fig. \ref{figure5}b). The BLS intensity shows a highly asymmetric distribution with respect to the signal line (Fig. \ref{figure5}c-f), in agreement with modeling results. We note (BLS intensity for $x < 0$)/(BLS intensity for $x > 0$) $\gg$ 1 in Fig. \ref{figure5}c, corresponding to giant lateral spin-wave asymmetry. The change in the asymmetry depending on the equilibrium magnetization direction is also consistent with modeling results. These results thus confirm the formation of laterally localized edge mode in cubic anisotropic materials, which is tunable by changing the equilibrium magnetization direction.

It is worthwhile comparing the laterally localized edge mode of cubic-anisotropy-based spin-wave devices with gate-voltage-controlled conducting channel of traditional three-terminal field-effect transistors (FETs). A similarity is that in the spin-wave devices, the edge channel is controlled by the equilibrium magnetization direction (which can be set by an external field), whereas in FETs, the conducting channel is controlled by a gate voltage. Therefore, the equilibrium magnetization direction of cubic-anisotropy-based spin-wave devices serves as a gate voltage of FETs. This similarity makes it possible to mimic the functionalities of three-terminal FETs with three-terminal cubic-anisotropy-based spin-wave devices, by controlling the equilibrium magnetization direction and measuring the induced voltage with an antenna placed at an edge (see Fig. \ref{figure4}d). Because of the giant asymmetry of edge modes, the on/off ratio of spin-waves [i.e., on (off) $\equiv$ spin-wave-induced voltage measured at an edge-antenna for $\phi=\pi/4$ ($\phi=3\pi/4$)] becomes very large.

There is also an important difference. Traditional FETs are volatile because the conducting channel is closed when the gate voltage is turned off. In contrast, the cubic-anisotropy-based spin-wave devices are {\it nonvolatile} because the spin-wave edge channel is maintained by the easy-axis anisotropy, even after removing an external field. In this respect, the cubic-anisotropy-based spin-wave devices can serve as three-terminal nonvolatile logic gates. In supplementary note 3, we propose proof-of-principle spin-wave logic gates performing Boolean functions of NOT, PASS, AND, NAND, OR, NOR, XOR, and XNOR. We note however that the proposed gates have inputs (spin-waves with a magnetic field) and outputs (only spin-waves) that are different so that they require additional spin-wave-to-current converters, which are definitely detrimental for practical use. One may combine spin-wave logic with spin-transfer torque~\cite{Demidov2016} or electric-field magnetization switching technique~\cite{Chiba} to remove or at least simplify this additional part.

\section{Conclusion}
\vspace{-3mm}
\noindent
We have demonstrated that the cubic anisotropic material is a promising candidate for coherent magnonic devices by virtue of largely enhanced spin-wave properties and laterally localized edge modes. The enhanced spin-wave properties will greatly improve the signal-to-noise ratio of magnonic devices. Up to now, conventional spin-wave nonreciprocity has been employed to generate a $\pi$ phase-shifted wave~\cite{Sekiguchi2} and different spin-wave amplitude~\cite{Jamali,Demidov,Sekiguchi2}, providing plenty of magnonic functionalities. The lateral asymmetry reported here will add a new functionality, three-terminal nonvolatile spin-wave logic function.


\section{Methods}
\noindent {\bf Sample preparation} The devices are deposited using magnetron sputtering with a base pressure of less than 7$\times$10$^{-7}$ Pa on MgO(001) single-crystal substrates. The high quality epitaxial Fe layer is fabricated by the multilayer stack of Cr(40)/Fe(25)/Mg(0.45)/Mg$_{19}$Al$_{81}$(1.2)/oxidation (thickness in nanometers)~\cite{Sukegawa}. The full-width half-maximum of the rocking curve for (002)-plane is reached to be 0.204$^{\circ}$ and the surface roughness is less than  $R_{a} \le $ 0.11 nm. The saturation magnetization $M_\text{s}=(1.6 \pm 0.1) \times 10^6$ A/m of epitaxial Fe films are determined by a vibrating sample magnetometer.
\\

\noindent {\bf Time-domain propagating spin-wave spectroscopy}
Spin-waves are excited and detected with a pair of asymmetric coplanar strip (ACPS) transmission lines. The ACPSs are designed to have the 50 $\Omega$ characteristic impedance, and patterned by electron beam lithography followed by a lift-off of Ti(5 nm)/Au(200 nm). A voltage pulse injected into the excitation ACPS generates a excitation magnetic field, which in turn excites a spin-wave packet. Propagating spin-wave induces a voltage on the detection ACPS connected to a 20 GHz sampling oscilloscope.
\\

\noindent {\bf Micromagnetic simulation}
Micromagnetic simulations are performed by numerically solving the Landau-Lifshitz-Gilbert equation, given as $\partial \textbf{m}/\partial t=-\gamma_{\rm g} \mu_0 \textbf{m} \times \textbf{H}_\text{eff}+\alpha_{\rm G} \textbf{m} \times \partial \textbf{m}/\partial t$ where $\textbf{m}$ is the unit vector along the magnetization, $\textbf{H}_\text{eff}$ is the effective magnetic field including the exchange, magnetostatic, and external fields, and $\alpha_{\rm G}$ is the Gilbert damping. The following parameters are used: The dimensions of the Fe film = 120 $\mu$m $\times$ 120 $\mu$m $\times$ 25 nm, the dimension of unit the cell = 500 nm $\times$ 500 nm $\times$ 25 nm, $\alpha_{\rm G}$ = 0.002, the exchange stiffness constant $A_\text{ex}$ = 1.3 $\times 10^{-11}$ J/m, the cubic anisotropy $K_\text{c}$ = 5.48 $\times 10^4$ J/m$^3$, and the saturation magnetization $M_\text{s}=1.66 \times 10^6$ A/m.
\\

\noindent {\bf BLS microscopy}
The spin-wave intensity is detected by BLS microscopy, which is based on the inelastic scattering of photons and magnons. A continuous, single-mode 532 nm solid-state laser is focused on the sample using a high numerical aperture microscope lens. The spatial resolution is 250 nm. The frequency shift of the inelastically scattered light is analyzed using a tandem (six-pass) Fabry-Perot interferometer. The sample position is monitored with a CCD camera and an active stabilization algorithm allows for controlling the sample position with respect to the laser focus with a precision better than 20 nm.


\section*{Acknowledgments}
        \vspace*{-0.5cm}
This work was supported by the Japan Science and Technology Agency, Precursory Research for Embryonic Science and Technology (JST-PRESTO). K. S. also acknowledges Grants-in-Aid for Scientific Research (25706004, 16H02098) from the Ministry of Education, Culture, Sports, Science and Technology, Japan. K. S. acknowledges D. Chiba and N. Ishida for valuable contributions. K.J.L. acknowledges the National Research Foundation of Korea (NRF) grant funded by the Korea government (MSIP) (2015M3D1A1070465, 2017R1A2B2006119) and Korea University Future Research Grant.

\section*{Author contributions}
        \vspace*{-0.5cm}
K.S. and K.-J.L. planned the experiment. K.S., H.S., and N.S. designed and prepared the samples.
K.S. and N.S. performed the time-domain spin-wave spectroscopy and the BLS microscopy.
S.W.L., S.O., R.D.M., and K.-J.L. performed analytical and numerical calculations to verify the effect of cubic magnetic anisotropy on spin-wave properties. K.S., S.W.L., R.D.M., and K.-J.L. wrote the paper. All authors discussed the results.

\section*{Additional information}
        \vspace*{-0.5cm}
Reprints and permissions information is available online at www.nature.com/reprints. Correspondence should be addressed to K.-J.L.

\section*{Competing financial interests}
        \vspace*{-0.5cm}
The authors declare no competing financial interests.

\clearpage

\begin{figure*}[thp]
\begin{center}
\psfig{file=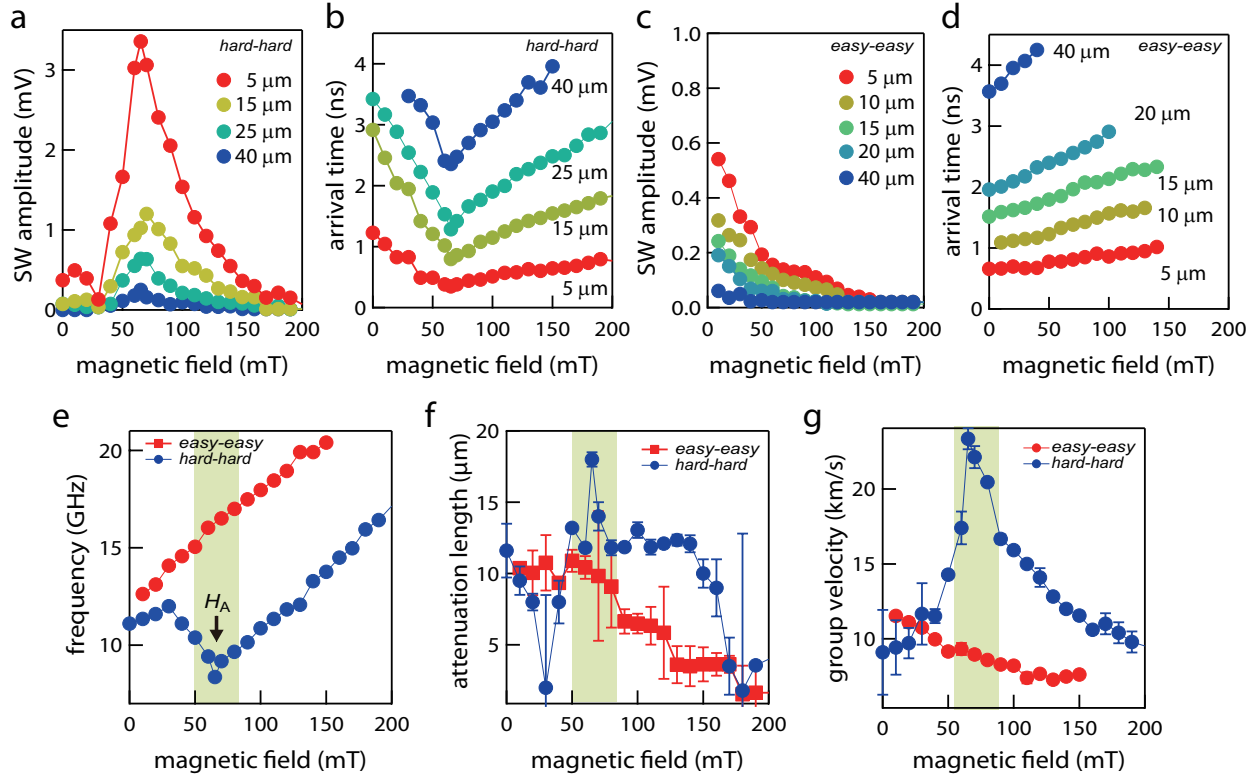,width=1.0\columnwidth}
   \caption{\label{figure3}{\bf Spin-wave properties in epitaxial Fe waveguide.} Magnetic-field dependences of {\bf a}, spin-wave amplitude and {\bf b}, arrival time of spin-wave packet for the \textit{hard-hard} case. {\bf c}, Spin-wave amplitude and {\bf d}, arrival time of spin-wave packet for the \textit{easy-easy} case. {\bf e}, The spin-wave frequencies deduced by FFT of the time-resolved waveforms. {\bf f}, Deduced attenuation length and {\bf g}, group velocity as a function of the magnetic field.  Where visible, error bars represent single standard deviations.  Otherwise, uncertainties are smaller than the symbols.} \label{figure3}
\end{center}
\end{figure*}

\begin{figure*}[bthp]
\begin{center}
   \includegraphics[width=20 cm,trim={0 0 0 3cm},clip]{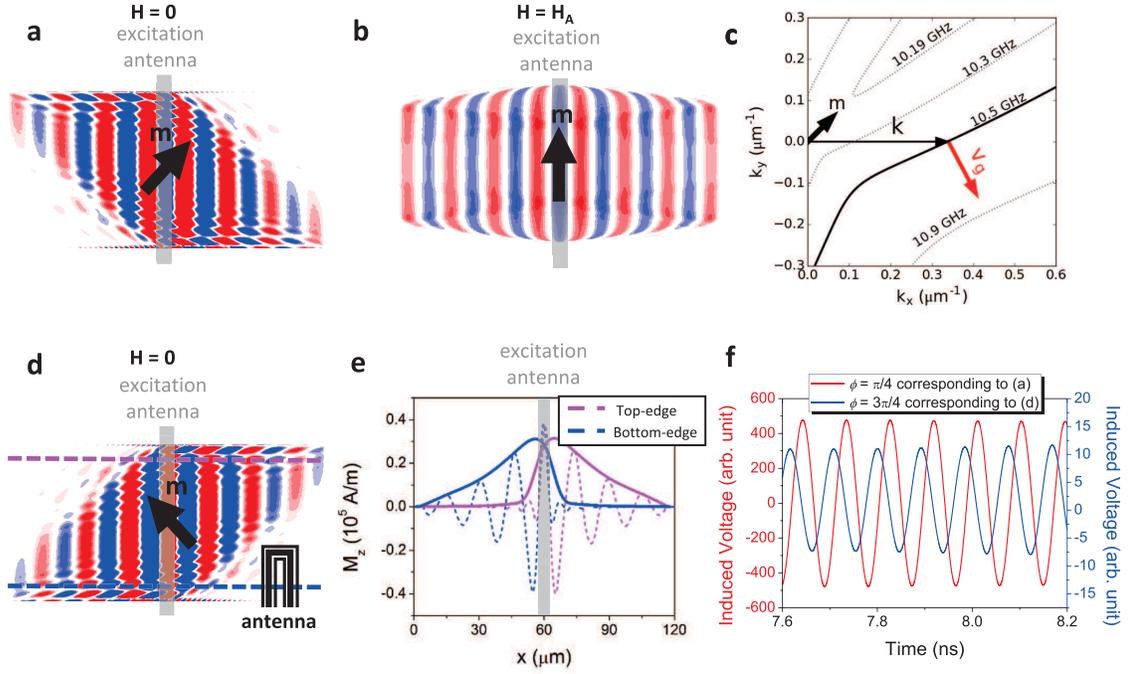}
        \vspace*{-2.5cm}
   \caption{\label{figure4}{\bf Two-dimensional micromagnetic simulations for laterally localized edge modes.} Snapshots of propagating spin-waves for {\bf a}, $H=0$ ($\phi=\pi/4$, $f=11.2$ GHz), {\bf b}, $H=H_{\rm A}$ ($\phi=\pi/2$, $f=9.6$ GHz), and {\bf d}, $H=0$ ($\phi=3\pi/4$, $f=11.2$ GHz). The direction of magnetization ($\bf{m}$) is depicted by a black arrow. {\bf c}, Contour plot of the spin wave dispersion relation at $H$ = 0 ($\phi=\pi/4$). The contour at the excitation frequency of $10.5$ GHz is highlighted. {\bf e}, Spin-wave amplitude profile along the top-edge line and bottom-edge line depicted in {\bf d}, showing the lateral asymmetry. {\bf f}, Induced voltages generated by the laterally asymmetric spin-waves (the detection antenna is assumed to be at 40 $\mu$m from the excitation antenna). The amplitude ratio between two cases is about 40.}
      \label{figure4}
\end{center}
\end{figure*}

\begin{figure*}[bthp]
\begin{center}
   \includegraphics[width=16 cm,clip]{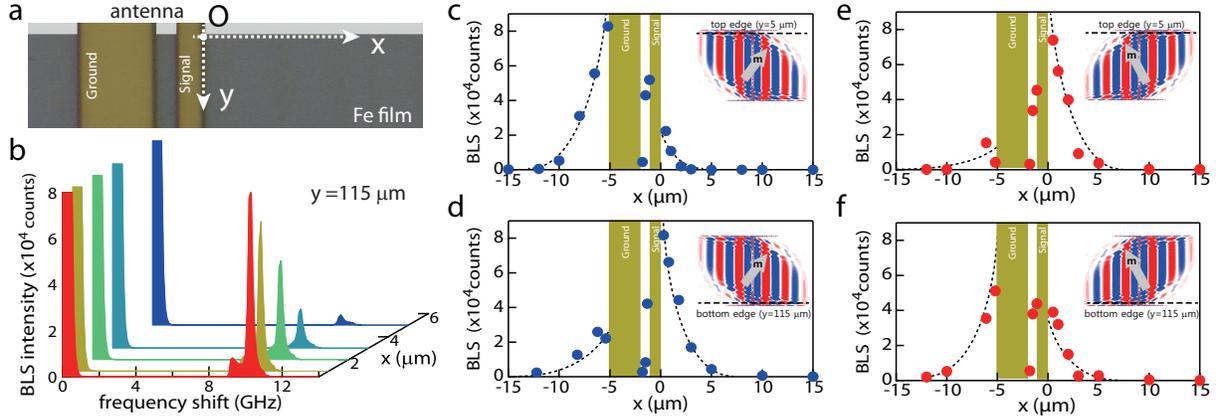}
   \nobreak
   \caption{\label{figure5}{\bf BLS detection of edge spin-wave mode.} Spatial distribution of magnon density in the remanent state ($H=0$). {\bf a}, BLS spectra are obtained at the edges $y= $5 and 115 $\mu$m. {\bf b}, Typical BLS spectra at the bottom edge (y = 115 $\mu$m) measured with the frequency window 9$< f <$12 GHz, containing a sharp peak centered at excitation frequency 10.4 GHz. Asymmetric distribution of magnon density at {\bf c}, the top edge (y =5 $\mu$m) with $\phi=\pi/4$, {\bf d}, bottom edge (y =115 $\mu$m) with $\phi=\pi/4$, {\bf e}, top edge (y =5 $\mu$m) with $\phi=3\pi/4$, and {\bf f}, bottom edge (y =115 $\mu$m) with $\phi=3\pi/4$. Insets show corresponding micromagnetic simulation results.}
      \label{figure5}
\end{center}
\end{figure*}

\end{document}